\documentstyle[amsbsy,amstex,aps,epsf,multicol]{revtex}
\def\eqref#1{Eq.~(\ref{#1})}

\def\phi{\varphi}

\def\({\left(}
\def\){\right)}
\def\[{\left[}
\def\]{\right]}
\def\<{\left\langle}
\def\>{\right\rangle}
\def\<{\left\langle}
\def\>{\right\rangle}

\def\bea{\begin{eqnarray}}
\def\eea{\end{eqnarray}}

\makeatletter

\title{On magnetic field generation in Kolmogorov turbulence}
\author{Stanislav Boldyrev${}^1$ and Fausto Cattaneo${}^2$ \\
{\em ${}^1$Department of Astronomy and Astrophysics,
 University of Chicago, 
5640 S. Ellis Ave, 
Chicago, IL 60637}\\ 
{\em ${}^2$Department of Mathematics,
 University of Chicago, 
5734 S. University, 
Chicago, IL 60637}\\
}

\date{\today}

\begin{document}

\input psfig.sty

%\doublespacing

\maketitle

\begin{abstract}
We analyze the initial, kinematic stage of magnetic field evolution 
in an isotropic and homogeneous turbulent conducting fluid 
with a rough velocity field, 
$v(l)\sim l^{\alpha}$, $\alpha<1$. 
We propose that in  the limit of small magnetic Prandtl number, i.e. 
when ohmic resistivity is much larger than viscosity, 
the smaller the roughness exponent $\alpha$,  
the larger the magnetic Reynolds number that is needed to excite 
magnetic fluctuations. This implies that numerical or experimental 
investigations 
of magnetohydrodynamic turbulence with small Prandtl numbers 
need to achieve extremely high resolution in order to  
describe magnetic phenomena adequately. 

% PACS numbers: 95.30.Qd, 95.30.Lz, 47.65.+a

\end{abstract}

\begin{multicols}{2}

{\bf 1.} {\em Introduction.} 
In a turbulent highly conducting fluid, magnetic fields may be amplified 
since the field lines are generally stretched by randomly moving fluid 
elements in which these lines are frozen~\cite{landau}. 
Such a mechanism of turbulent 
dynamo is expected to work in a variety of astrophysical systems (galaxy 
clusters, interstellar medium, stars, planets), 
is confirmed numerically, and is 
consistent with simple analytical models. 

Valuable insight can be gained from considering the 
so-called kinematic stage of the dynamo, when magnetic field is 
amplified from an initially weak `seed' field. As the simplest example, 
assume that the resistive 
scale is smaller than or of the same order as the viscous scale of the fluid. 
In the turbulent Kolmogorov velocity field, the smallest eddies  
have the highest shearing rate given by $v(l)/l\sim l^{-2/3}$, 
where $l$ is the size of the eddy, and $v(l)\sim l^{1/3}$ 
is its turn-over velocity. Therefore, 
at this stage the magnetic field grows predominantly on small scales.  
Fig.~\ref{first_stage} shows schematically the initial stage of 
the evolution of the magnetic spectrum, see, e.g.,~\cite{kulsrud}.

The magnetic energy collapses toward small, resistive scale 
during this initial evolution, until the field is strong enough to affect 
the dynamics of fluid through the Lorentz force.  Such behavior 
is the evidence of small-scale turbulent dynamo; it is firmly 
established in numerical experiments, and can be derived 
analytically~\cite{kulsrud,batchelor,kazantsev,boldyrev1,gruzinov,falkovich,lazarian}. 
Since in this example, the resistive 
scale is smaller than the viscous one, the 
dynamo is essentially governed by a smooth, viscous-scale velocity field.

To study the growth of magnetic energy on larger scales we have  
to understand how the magnetic field is generated in the inertial  
interval of the turbulence (the interval of scales much 
smaller than the external scale where the turbulence is excited, 
and much larger than the viscous scale where the turbulent energy 
is dissipated).  This question is non-trivial since  
in this interval the velocity field is not smooth, i.e., 
$v(l)\sim l^{\alpha}$, with $\alpha<1$. This situation is
especially relevant for the case of small magnetic Prandtl 
number ($\mbox{Pm}=\nu/\eta$, where $\nu$ is fluid viscosity 
and $\eta$ is resistivity), where the  magnetic 
energy is concentrated mostly in the inertial interval of 
the velocity field.

The question was first addressed by Batchelor~\cite{batchelor} who 
used the analogy between magnetic field lines and fluid vorticity lines 
to conclude that when $\nu\leq \eta$, magnetic energy is not amplified. 
This analogy was criticized in~\cite{kraichnan_nagarajan} because the 
magnetic field can have arbitrary initial conditions, while the vorticity 
field is related to the velocity field. It was further 
argued in~\cite{vainshtein} that magnetic field line 
stretching generally dominates resistive reconnection, thus making  
dynamo possible for $\nu \leq \eta$. Direct numerical 
simulations of MHD turbulence with $\mbox{Pm}=1$ confirm that 
week magnetic fluctuations are generally amplified by the Navier-Stokes 
velocity field except 
for a special case when the initial magnetic field is close to the vorticity 
field~\cite{tsinober,ohkitani}.
{
\columnwidth=3.4in
\begin{figure} [tbp]
\centerline{\psfig{file=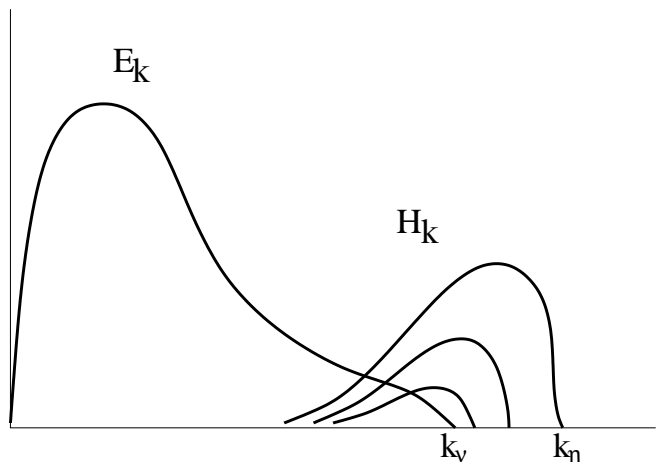,width=3.4in,angle=0}}
\vskip3mm
\caption{Sketch of the initial stage of magnetic field 
amplification by turbulent 
velocity field. $E_k$ is the spectrum of the velocity field in $k$ space, 
$H_k$ is the spectrum of the magnetic field. The viscous cut-off $k_{\nu}$ is 
larger than the resistive cut-off $k_{\eta}$. Magnetic field energy 
is shown for three consecutive times, it grows and gets 
accumulated at the resistive scale.
}
\label{first_stage}
\end{figure}
}

Recently, there appeared the number of 
numerical simulations of MHD turbulence with small magnetic Prandtl 
numbers, where magnetic fluctuations were not 
amplified~\cite{cattaneo,christensen,cowley,yousef}, which revived 
the claims that dynamo does not exist in the Kolmogorov turbulence 
with $\mbox{Pm} \ll 1$, see e.g., \cite{cowley,yousef}.
However, astrophysical observations show that magnetic fields are 
generated by turbulent motion rather effectively in planets 
and stars where magnetic Prandtl 
numbers are small (e.g., in the 
geo-dynamo, $\mbox{Pm}\sim 10^{-5}-10^{-6}$, 
in the solar photosphere, $\mbox{Pm}\sim 10^{-7}$). 
This apparent contradiction motivated our interest to the problem.

In this paper we argue that 
dynamo always exists in a rough velocity field, such as the 
Kolmogorov field. We however find that the magnetic Reynolds 
number, $\mbox{Rm}=v(L_0)L_0/\eta$, and the numerical resolution 
required to generate magnetic field strongly 
depend on the roughness exponent of the 
velocity field, $\alpha$; the rougher the velocity field, 
the larger the required resolution. This result explains why 
dynamo action  
is hard to achieve in experiments  
with small magnetic Prandtl numbers, 
while it is easily achieved when the magnetic Prandtl number is large, 
$\mbox{Pm}\geq 1$. In the latter case the velocity field is 
effectively smooth at the resistive scales, where the  
magnetic energy is concentrated. 

Our analysis is based on the 
so-called Kazantsev model of kinematic dynamo, which 
is exactly solvable and  which allows us 
to change the velocity roughness exponent, $\alpha$. 

{\bf 2}. {\em The Kazantsev model for a rough velocity field}. 
Consider the induction equation for the magnetic field:
\begin{eqnarray}
{\partial {B^i}}/{\partial t}+v^j\partial B^i/\partial x^j
-B^j\partial v^i/\partial x^j=\eta \Delta B^i,
\label{induction}
\end{eqnarray}
where $B^i({\bf x}, t)$ is the magnetic field, $v^i({\bf x},t)$ 
is the fluid velocity, $\eta$ is small ohmic resistivity, and we 
sum over the repeated vector indices. At the 
initial stage of evolution, the weak magnetic field is passively  
advected by the fluid. Therefore, one can use simplifying 
assumptions about the fluid velocity to make the problem manageable. 
Kazantsev~\cite{kazantsev} and Kraichnan~\cite{kraichnan} introduced 
the model based on the Gaussian, short-time correlated 
velocity field, with zero mean and the covariance
\begin{eqnarray} 
\langle v^i({\bf x}, t)v^j({\bf x}', t') \rangle
=\kappa^{ij}({\bf x}-{\bf x'})\delta(t-t').
\label{correlator}
\end{eqnarray}
This model is a valuable tool for the analytical investigation of 
kinematic dynamo;  
direct numerical simulations reveal that a purely 
Gaussian velocity field amplifies small magnetic fluctuations in a similar 
manner as the true Navier-Stokes field~\cite{tsinober}.

Assuming isotropy and homogeneity, the velocity correlation 
function has the form
\begin{eqnarray}
\kappa^{ij}({\bf r})=\kappa_N(r)\left(\delta^{ij}-\frac{r^ir^j}{r^2} 
\right)+\kappa_L(r)\frac{r^ir^j}{r^2},
\label{kappa}
\end{eqnarray}
where ${\bf r}={\bf x}-{\bf x'}$. If we further assume that the  
velocity field is incompressible, we have 
$\kappa_N=\kappa_L+(r\kappa_L^\prime)/2$, and  
velocity statistics can be characterized by only one scalar function, 
$\kappa_L(r)$. 

The model defined by (\ref{induction}), (\ref{correlator}), 
and (\ref{kappa}) allows one to write a closed equation for the 
correlation function of the magnetic field,
\begin{eqnarray}
\langle B^i({\bf x}, t)B^j({\bf x}', t) \rangle=H^{ij}({\bf x}-{\bf x}', t),
\label{bcorrelator}
\end{eqnarray}
where, analogously to (\ref{kappa}), the $H^{ij}$ function can be 
represented as
\begin{eqnarray}
H^{ij}=H_N(r,t)\left(\delta^{ij}-\frac{r^ir^j}{r^2} 
\right)+H_L(r,t)\frac{r^ir^j}{r^2},
\end{eqnarray}
furthermore, the 
condition $\nabla\cdot {\bf B}=0$ gives $H_N=H_L+(rH_L^\prime)/2$.
We will characterize the magnetic field correlator by the function $H_L(r,t)$. 
The equation for this function can be found by differentiating 
(\ref{bcorrelator}) with respect to time, and by using 
Eqs. (\ref{induction})-(\ref{kappa}). A rather tedious but essentially 
straightforward  calculation gives:
\begin{eqnarray}
{\partial_t H}_L=\kappa H_L^{\prime \prime} + \left(\frac{4}{r}\kappa+\kappa' \right)H_L^{\prime}
+\left( \kappa''+\frac{4}{r}\kappa' \right)H_L,
\label{kequation}
\end{eqnarray}
where primes denote the 
derivatives with respect to~$r$ and we have introduced 
the `renormalized' velocity correlation 
function $\kappa(r)=2\eta+\kappa_L(0)-\kappa_L(r)$. 
Equation~(\ref{kequation}) was originally derived 
by Kazantsev~\cite{kazantsev}, and can be rewritten in a different form 
that formally coincides with the Shr\"odinger equation in the imaginary time. 
Effecting the change of variable, 
$H_L=\psi(r,t) r^{-2}\kappa(r)^{-1/2}$, one obtains:
\begin{eqnarray}
\partial_t \psi=\kappa(r)\psi''-V(r)\psi,
\label{sequation}
\end{eqnarray}
which describes the wave function of a quantum particle with the variable 
mass, $m(r)=1/[2\kappa(r)]$, in a one-dimensional potential~($r>0$):
\begin{eqnarray}
V(r)=\frac{2}{r^2}\kappa(r)-\frac{1}{2}\kappa''(r)-\frac{2}{r}\kappa'(r)-
\frac{(\kappa'(r))^2}{4\kappa(r)}.
\end{eqnarray}

This equation can be investigated for different choices 
of~$\kappa(r)$, however, we restrict ourselves to the inertial 
interval of the turbulence, where the velocity correlator has  
power-law asymptotics, $\kappa(r)\propto r^{\beta}$. The exponent 
$\beta$ can be found from the scaling of turbulent diffusivity, 
$D\sim v(r) r\sim r^{1+\alpha}$. Indeed, in the derivation of 
Eq.~(\ref{kequation}) we used the 
integral $D=\int_0^{\infty} \langle  [v(x,t)- v(x',t)][v(x,t')- v(x',t')] 
\rangle d(t-t')= \kappa(r)$, which is the turbulent 
diffusivity~\cite{vainshtein}. 
Comparing the two expressions we obtain $\beta=1+\alpha$. 

The 
Shr\"odinger equation~(\ref{sequation}) has the effective 
potential $U_{eff}(r)=V(r)/\kappa(r)=A(\beta)/r^2$,  
where $A(\beta)=2-3\beta/2-3\beta^2/4 $. 
When $A(\beta)< -1/4 $, the quantum particle falls toward 
the origin {\cite{landau_qm}} and its wave function is therefore 
concentrated at the smallest, 
resistive scale. This behavior is the manifestation of the dynamo 
mechanism that we discussed in the introduction. 
We obtain that $A(1+\alpha)<-1/4$ for any roughness 
exponent of the velocity field, $0<\alpha <1$, therefore, dynamo is 
always possible; the same conclusion was made in~\cite{vainshtein}. 

{\bf 3.} {\em Magnetic field correlator and dynamo growth rates.} 
To find the wave function (the magnetic field correlator) 
we need to specify the boundary conditions.  
The small-scale regularization is naturally given 
by ohmic resistivity. For scales much smaller than 
the correlation scale of the velocity field, the $\kappa$ function 
can be expanded as $\kappa(r)\simeq \kappa_L(0)(2\eta+r^{1+\alpha})$, 
which corresponds to the limit of infinitely small Prandtl number. 
In this formula and below we use the dimensionless 
variables: $\eta$ is measured 
in the units of the large-scale turbulent diffusivity,~$\kappa_L(0)$, 
while $r$ 
is in the units of the integral scale,~$L_0$.

The boundary condition at the origin follows from finiteness of 
magnetic energy, $\lim\limits_{r\to 0}H_L(r,t)=H_0(t)<\infty$. The boundary 
condition at large scales follows from the absence of the  
mean magnetic field, $\lim\limits_{r\to \infty}H_L(r,t)=0$. We will see in a 
moment that at the kinematic stage magnetic energy is concentrated 
at the resistive scale, and the corresponding wave function decays 
exponentially fast in the inertial interval, $r\gg r_{\eta}$, 
so it is indeed independent of  the 
large-scale properties of the velocity field, as we expected.  

Problem (\ref{kequation}) can be cast into the Sturm--Liouville form  
by changing $H_L(r,t)=h(r,t)/r^2$, which teaches us that its solution 
can be expanded in the eigenfunctions of the corresponding Sturm--Liouville 
operator. The maximum growth rate is therefore given by the largest 
eigenvalue of the operator on the right-hand side 
of Eqs.~(\ref{kequation}) or (\ref{sequation}). The corresponding 
wave function is the ground state of the potential~$U_{eff}(r)$. 
Thus, following Kazantsev~\cite{kazantsev}, we look for the solution 
of (\ref{sequation}) in the 
form $\psi(r,t)=\phi(r)\exp{(\lambda t)}$. 

In the inertial interval, $2\eta \ll r^{\beta}$ , the equation 
for the $\phi$ function reads 
\begin{eqnarray}
-\phi^{''}+ \left[\frac{-3\beta^2-6\beta+8}{4 r^2}+
\frac{\lambda}{r^\beta} \right]\phi=0,
\label{inertialequation}
\end{eqnarray}
where $\lambda$ is dimensionless and is measured in the 
units of $\kappa_L(0)/L_0^2$. For small $\lambda$, the potential 
is dominated by the first term in the square brackets, and one 
can easily check that the corresponding wave function oscillates 
in the inertial interval. However, the ground-state wave function cannot 
oscillate, therefore the ground-state growth rate, $\lambda$, will 
be such that the second term in the square brackets 
of (\ref{inertialequation}) dominate the first one in the 
inertial interval. At the resistive scale, $r_{\eta}= (2\eta)^{1/\beta}$, 
both terms should be of the same order, therefore, 
$\lambda\sim \eta^{(\beta-2)/\beta}$.

The wave function corresponding to the growing solution, $\lambda>0$, decays 
exponentially fast for $r \gg \eta^{1/\beta}$,   
$\phi \propto \exp\left(-\frac{2 \sqrt{\lambda} r^{(2-\beta)/2}}
{2-\beta}\right)$. 
This function is concentrated at the resistive scale and its growth rate 
is of the order of the eddy turn-over time at this scale; in 
agreement with our qualitative discussion in the introduction. When 
one changes the magnetic Prandtl number (by changing viscosity, 
for instance) the effective roughness of the velocity field at the 
resistive scale changes. We therefore suggest that the effect of 
different Prandtl numbers 
can be studied in terms of Eq.~(\ref{inertialequation}) with 
different roughness exponents. By this analogy, 
the limit of the smooth velocity field, $\beta=2$,  corresponds 
to large $\mbox{Pm}$, while the other limit of the Kolmogorov-scaled 
velocity field, $\beta=4/3$, corresponds to small $\mbox{Pm}$.  

We observe that the rougher the velocity field, the broader  
the wave function compared to the 
resistive scale, 
%$r_0/r_{\eta} \propto (2-\beta)^{2/(2-\beta)}$,  
and, therefore, the larger the magnetic Reynolds number necessary to 
generate magnetic fluctuations.
To estimate the critical resolution for the dynamo onset, 
we solved Eq.~(\ref{kequation}) numerically with the 
large-scale boundary condition,~$H_L(\ell)=0$. 
For given $\beta$ and $\eta$, we increased 
the `system size',~$\ell$, until $H_L$ started 
to grow and we thus found the critical value~$\ell=\ell_c$. 
An analogous calculation could be done by fixing $\ell$ and changing $\eta$. 

To characterize the inertial range, we introduce the 
resolution parameter, $R=\ell_c/r_{\eta}=\ell_c/(2\eta)^{1/\beta}$; this 
parameter is universal, i.e., it is independent of large and small-scale 
properties of the velocity field.
We obtain that when the velocity roughness increases 
from~$\beta=2.0$ 
to~$\beta=1.3$, the corresponding  
resolution parameter increases from $R\approx 3.8$ to 
 $R\approx 29$, see Fig.~(\ref{resolution}).  
This means that if we simulated the random 
equation~(\ref{induction}) directly, 
the required numerical resolution should 
increase by about an order of magnitude in each spatial direction 
as we go from the smooth to the Kolmogorov-scaled velocity field. 
{
\columnwidth=3.4in
\begin{figure} [tbp]
\centerline{\psfig{file=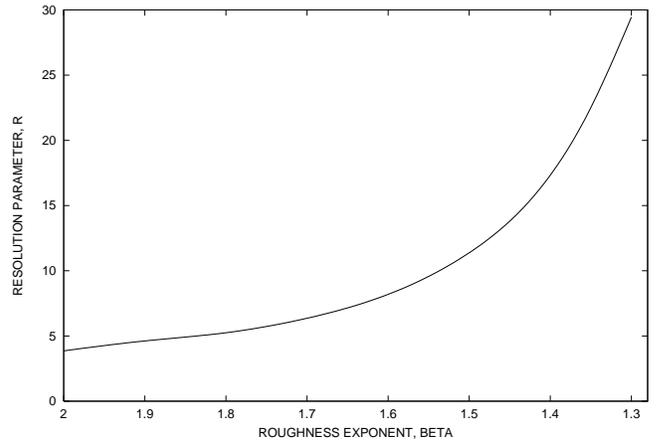,width=3.4in,angle=-90}}
\vskip3mm
\caption{The numerically found resolution parameter, $R=\ell_c/r_{\eta}$, as a 
function of the velocity roughness exponent,~$\beta$.
}
\label{resolution}
\end{figure}
}

The subresistive scales, $r \ll \eta^{1/\beta}$, can be easily  
described using Eq.~(\ref{kequation}). By direct substitution, one 
can check that the magnetic field correlator can be expanded as 
$H_L(r,t)= H_0[1-r^{\beta}/(2\eta)+... ]\exp(\lambda t)$, which 
implies that the spectrum of the magnetic field 
decays as $H_k \sim k^2|B_k|^2\propto k^{-1-\beta}$ in the subresistive 
region, $\eta^{-1/\beta}\ll k \ll \nu^{-1/\beta}$. 
With such a spectrum, the rate of magnetic energy dissipation at the 
viscous scale is bigger than the rate at the resistive scale by 
the factor 
$k_{\eta}^5 |B_{k_{\eta}}|^2/k_{\nu}^5|B_{k_{\nu}}|^2
\sim \mbox{Pm}^{-(2-\beta)/\beta}$.
 Thus, the effective resistivity may  
 be `anomalously large' in turbulent flows with small Prandtl numbers. 

Strictly speaking, the last result requires short-time correlated 
small-scale 
eddies and is not applicable in 
the Kolmogorov turbulence where velocity is correlated at the eddy turn-over 
time, $\tau_{corr}(l) \sim l/v(l)$. At the subresistive 
scales, $l<l_{\eta}$, this correlation time   
is larger than the resistive relaxation time, $\tau_{\eta}\sim l^2/\eta$, 
while in the Kazantsev model 
the velocity correlation time is smaller than the resistive 
relaxation time. However, model (\ref{correlator}) 
is physically self-consistent. The large magnetic energy dissipation 
is balanced by the large energy transfer from the small-scale velocity 
eddies to the small scale magnetic field;  this is possible since 
the fluid energy is formally infinite in~(\ref{correlator}). 
In a practical situation where 
the velocity correlation time is not infinitely small, the 
asymptotic tail $H_k\sim k^{-1-\beta}$ will hold up to the scales where 
this correlation time is comparable to the resistive 
time, $k\sim 1/\sqrt{\eta \tau_{corr}}$. One can expect that for 
smaller scales, the magnetic energy spectrum will have a steeper decay, 
see, e.g,~\cite{kraichnan_nagarajan,golitsyn,moffatt}.

{\bf 4}. {\em Conclusions}.
We proposed that magnetic fluctuations are always generated in 
isotropic and homogeneous three-dimensional turbulence if the 
magnetic Reynolds number is large enough. The required critical  
magnetic Reynolds number sharply increases with 
velocity roughness, which explains 
the `no-dynamo' outcomes reported in numerical 
simulations with small Prandtl numbers, 
see e.g.,\cite{cattaneo,christensen,cowley,yousef}. 
We do not see  
any physical reason for the absence of dynamo in these numerical simulations 
rather than lack of resolution. Our results also 
suggest that obtaining small-scale dynamo will 
be a serious challenge for laboratory experiments, where 
the magnetic Prandtl number is small, 
$\mbox{Pm}\sim 10^{-5}$, while the magnetic Reynolds number is 
rather moderate, $\mbox{Rm}\sim 100$ 
\cite{gailitis1,gailitis,bourgoin,forest}.

As we mentioned in the introduction, the Batchelor analogy of magnetic 
field lines and vorticity lines would hold for $\nu=\eta$ 
if the initial magnetic 
field were proportional to the vorticity field. In this special case, 
magnetic energy would not be amplified.  
In the model we investigate there are infinitely many such 
special initial conditions; any magnetic correlator $H_L(r)$ lacking the 
growing eigenfunctions in its expansion will not be 
amplified. This, of course, does not mean that dynamo does not exist.

We also note that the Schr\"odinger-type 
equation of the form~(\ref{sequation}) was considered for the 
magnetic field correlator in the case of 
a general velocity field in \cite{vainshtein}. In this 
sense the applicability of the equation may not be restricted 
to the Gaussian, short-time correlated velocity field only. 
For example, we also investigated  
the modified Kazantsev model, with a more realistic,  
finite-time correlated velocity field. The velocity 
correlation time at a given wave number, $k$, was chosen to be 
of the order of the eddy turnover time at the corresponding scale. 
This model could not be solved analytically, however, its numerical 
integration gave qualitatively the same results 
as~(\ref{induction},\ref{correlator}). We will report these 
results elsewhere.

We would like to thank S. C. Cowley, R. Rosner, A. Schekochihin, 
and  S. I. Vainshtein for many useful discussions.
This work was supported by the NSF Center for magnetic 
self-organization in astrophysical and laboratory plasmas at 
the University of Chicago.  S.B. would like to thank the 
Aspen Center for Physics, where this work was initiated.

\end{multicols}

\end {document}